\documentclass[10pt,conference]{IEEEtran}
\IEEEoverridecommandlockouts
\usepackage{cite}
\usepackage{amssymb,amsfonts}
\usepackage[nosumlimits]{amsmath}
\usepackage{algorithmic}
\usepackage{textcomp}
\usepackage{hhline}
\def\BibTeX{{\rm B\kern-.05em{\sc i\kern-.025em b}\kern-.08em
    T\kern-.1667em\lower.7ex\hbox{E}\kern-.125emX}}
\usepackage{amsmath,graphicx}
\usepackage{multirow}
\usepackage{enumitem}
\usepackage[table]{xcolor}
\usepackage[caption=false, font={scriptsize}]{subfig}

\begin{document}

\title{A Machine Learning Approach for Denoising and Upsampling HRTFs\\
% {\footnotesize \textsuperscript{*}Note: Sub-titles are not captured for https://ieeexplore.ieee.org  and
\thanks{This research is supported by the SONICOM project (EU Horizon 2020
RIA grant agreement ID: 101017743).}
}

% \and
% \and
% \and
% \and
% \and
\author{
    \IEEEauthorblockN{Xuyi~Hu\IEEEauthorrefmark{1}, Jian~Li\IEEEauthorrefmark{1}, Lorenzo~Picinali\IEEEauthorrefmark{1} and Aidan~O.~T.~Hogg\IEEEauthorrefmark{2}\IEEEauthorrefmark{1}}
    \IEEEauthorblockA{\IEEEauthorrefmark{1}Audio Experience Design, Dyson School of Design Engineering, Imperial College London, UK}
    \IEEEauthorblockA{\IEEEauthorrefmark{2}Centre for Digital Music, School of Electronic Engineering and Computer Science, Queen Mary University of London, UK}
}

\maketitle

\begin{abstract}
The demand for realistic virtual immersive audio continues to grow, with Head-Related Transfer Functions (HRTFs) playing a key role. HRTFs capture how sound reaches our ears, reflecting unique anatomical features and enhancing spatial perception. It has been shown that personalized HRTFs improve localization accuracy, but their measurement remains time-consuming and requires a noise-free environment. Although machine learning has been shown to reduce the required measurement points and, thus, the measurement time, a controlled environment is still necessary. This paper proposes a method to address this constraint by presenting a novel technique that can upsample sparse, noisy HRTF measurements. The proposed approach combines an HRTF Denoisy U-Net for denoising and an Autoencoding Generative Adversarial Network (AE-GAN) for upsampling from three measurement points. The proposed method achieves a log-spectral distortion (LSD) error of 5.41 dB and a cosine similarity loss of 0.0070, demonstrating the method's effectiveness in HRTF upsampling.
\end{abstract}


\section{Introduction}
\label{sec:intro}
We live in an ever more digital world where the need to create realistic, immersive audio is becoming ever more essential. The implications of being able to create convincing immersive audio virtually are broad, not only in helping enhance augmented and virtual meetings or video games, but also playing an essential role in improving assistive technologies. These include but are not limited to hearing aids \cite{foerster2023reduced} and advancing speech intelligibility algorithms\cite{tan2023implementing}. 

One of the main challenges of immersive audio is adapting to individual listeners \cite{picinali2022system}. This individualization has resulted in a large amount of research focusing on Head-Related Transfer Functions (HRTFs). HRTFs describe how sound waves are modified as they propagate from a sound source to the listener's ears. This filtering is affected by a number of factors, including sound diffraction, reflection, and absorption by the listener's head and torso, as well as the resonances and shape of the listener's outer ears (pinnae). As a result, HRTFs contain both binaural cues, including interaural level differences (ILDs) and interaural time differences (ITDs), and monaural spectral cues, which are crucial for sound localization. Therefore, when the sound is appropriately filtered by the HRTF and presented at the entrance of the listener's ear canals, the listener should be able to perceive the sound originating from a specific source location \cite{blauert1985spatial}. This underscores the significant role HRTFs play in creating a realistic and immersive audio environment, particularly in applications such as virtual reality (VR)\cite{brahimaj2024enhancing, ramirez2024toward} and augmented reality (AR) \cite{privitera2023effect, privitera2023preliminary}.

Personalized HRTFs are closely tied to an individual’s anatomy, with each person possessing a unique HRTF. Utilizing non-individualized HRTFs in virtual simulations often leads to poor sound source localization performance \cite{stitt2019auditory,wenzel1993localization}. To ensure an optimal user experience, acquiring a personalized HRTF is crucial. One common approach involves acoustic measurements \cite{engel2023sonicom}, where sine sweeps are played from specific source locations, recorded at the listener's ears, and analyzed to extract impulse responses for generating the HRTF. However, this process requires specialized equipment and controlled environments, making it time-intensive \cite{li2020measurement}.

% An alternative is numerical calculation approaches \cite{tz2001boundary,kreuzer2009fast,harder2016framework}, which utilize anatomical structural information to compute individualized HRTFs. In addition, such anthropometric data can be utilized to choose the most suitable HRTF from publicly available HRTF databases \cite{geronazzo2019applying,zotkin2003hrtf}. Other selection criteria can be adopted, such as perceptual-based methods \cite{katz2012perceptually,romigh2014do}, or their combinations \cite{algazi2002approximating,brown1998structural}. Nevertheless, obtaining an accurate 3D representation of the listener's pinnae itself is a challenging problem, which may involve costly setup such as CT scans \cite{ziegelwanger2013calculation} and MRI \cite{greff2007round}.
To improve the efficiency and scalability of HRTF personalization, spatial up-sampling has been introduced to address the limitations of low-resolution HRTF data, which typically includes sparse measurements from limited directions. This technique generates high-resolution HRTFs by increasing measurement density, enhancing accuracy and coverage. Two common approaches are Barycentric interpolation \cite{hartung1999comparison,cuevas2019tunein} and spherical harmonic (SH) interpolation \cite{engel2022assessing,arend2021assessing}. Barycentric interpolation uses weighted averages of known points to estimate values in unmeasured areas, while SH interpolation projects the HRTF onto spherical basis functions for smooth spatial representation. These methods have significantly advanced HRTF up-sampling, enabling more accurate and individualized sound localization.

\begin{figure*}[tp]
\centering
\includegraphics[width=0.85\linewidth]{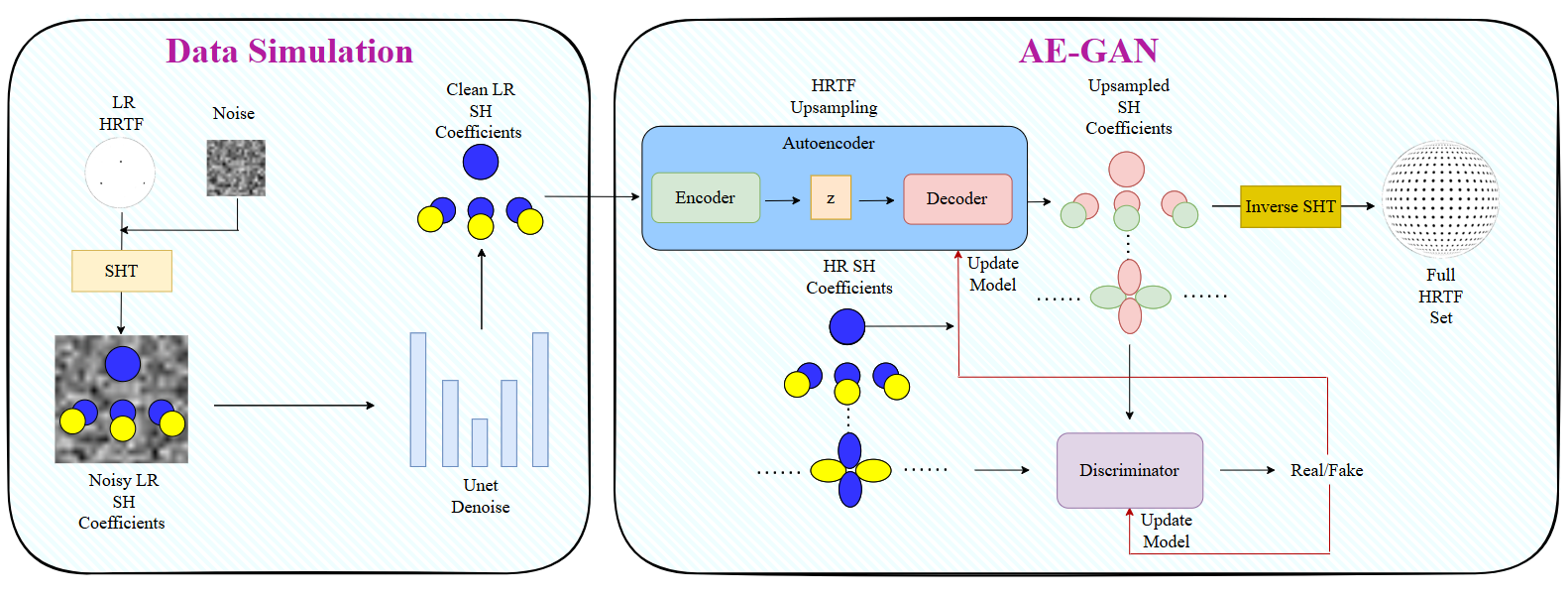}
\caption{\label{fig:U-GANet}HRTF-DUNet Flowchart. The left panel illustrates the data simulation pipeline, where noisy HRTF data is generated, segmented, and transformed using spherical harmonic analysis, resulting in low-resolution noisy coefficients stored in a dataset. The Denoisy U-Net then reconstructs clean SH coefficients from these inputs. The right panel presents the overall model framework. Red arrows indicate the AE-GAN training process, including the feedback loop for parameter updates, while black arrows represent the feedforward process through the model.}
\vspace{-0.3cm}
\end{figure*}

In recent years, machine learning (ML) methods have also emerged as promising approaches for HRTF personalization \cite{chen2019autoencoding,siripornpitak2022spatial,hogg2023exploring}. Techniques such as variational autoencoders (VAEs) \cite{lopez2018information} encode HRTFs into a latent space to reduce dimensionality while preserving key features, enabling reconstruction and upsampling of low-resolution HRTFs by filling in missing details. Generative adversarial networks (GANs) \cite{hogg2024hrtf, hu2024hrtf} use a generator to produce synthetic HRTFs and a discriminator to differentiate between real and synthetic data, learning data distributions to generate detailed high-resolution HRTFs. The SONICOM Listener Acoustic Personalization (LAP) Challenge 2024 demonstrated the superiority of ML techniques over traditional signal processing approaches \cite{hogg2025, geronazzotechnical
}.

Despite their advancements, these ML approaches still rely on clean and high-resolution data, necessitating noise-free environments for recording accurate HRTF measurements. To overcome this constraint, denoising (as well as upsampling) is needed, as real-world measurements are often degraded by background noise and room interactions, particularly in non-acoustically treated settings. Addressing these challenges is critical to improving accessibility, enabling accurate HRTF measurements, and expanding the adoption of immersive audio technologies and their applications. The contributions of this paper can, therefore, be broken down as follows:
\begin{enumerate}
    \item We enhance the AE-GAN approach from the authors' previous work\cite{hu2024hrtf} on HRTF upsampling.
    \item We employ an HRTF Denoisy U-Net for the task of denoising HRTFs measured in simulated noisy conditions. 
    \item We propose a novel end-to-end framework (HRTF-DUNet) and evaluate its performance against four baselines (AE-GAN without DUNet, Barycentric interpolation, SH interpolation, and HRTF selection) in terms of the LSD on the SONICOM HRTF dataset \cite{engel2023sonicom}.
\end{enumerate}

\section{METHOD}
\label{sec:method}
\subsection{HRTF Denoisy U-Net}
The proposed approach improves SH interpolation by using an AE-GAN to increase the SH order. Therefore, the noisy HRTF data first needs to be transformed into the SH domain using the Spherical Harmonic Transform (SHT) as part of a pre-processing step \cite{arend2021assessing}. The noisy, low-resolution SH coefficients are then passed to the HRTF Denoisy U-Net, which outputs the denoised low-resolution SH coefficients. 

\begin{figure*}[!tb]
\centering
\includegraphics[width=0.75\linewidth]{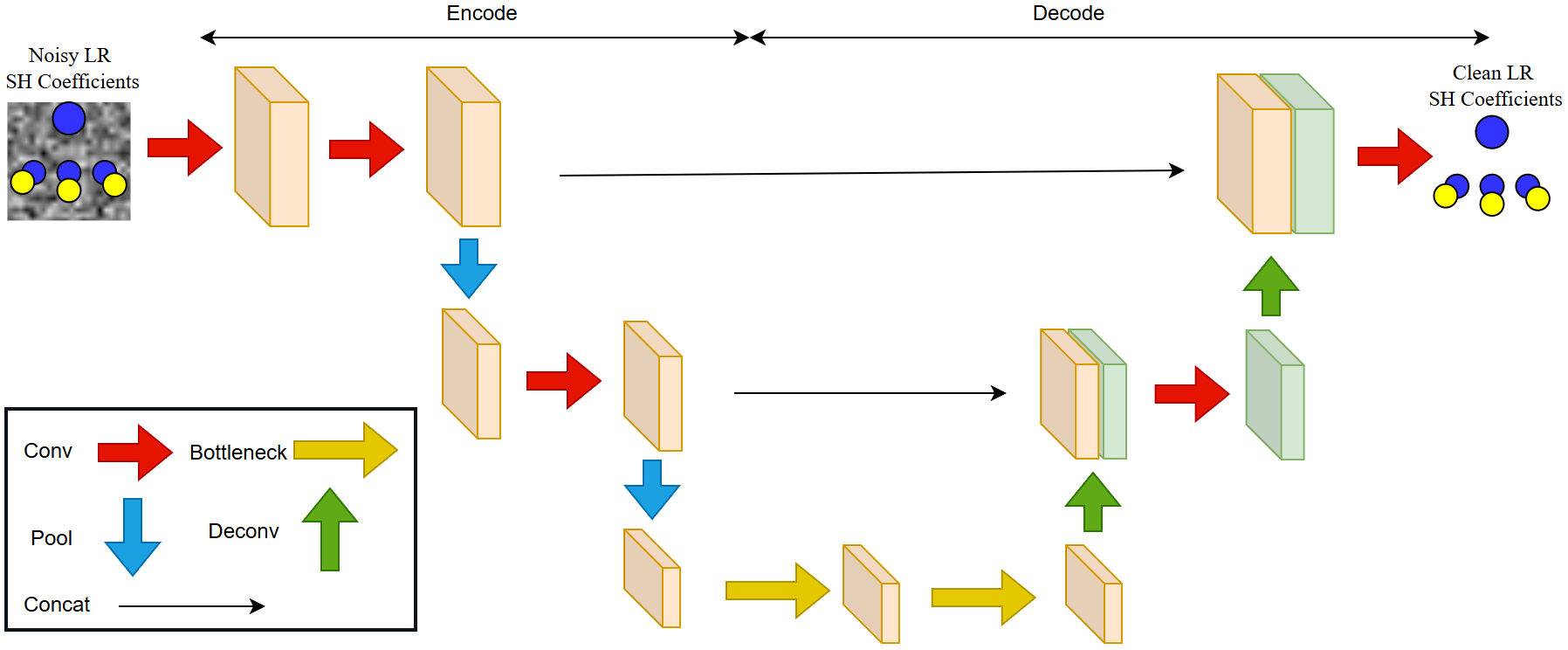}
\caption{\label{fig:U-Net}Scheme of the proposed HRTF Denoisy U-Net.}
\vspace{-0.4cm}
\end{figure*}

The Denoisy U-Net architecture is shown in Fig. \ref{fig:U-Net} and consists of an initial convolutional block that processes the noisy input SH coefficients. Each convolutional block in the network applies a 1D convolutional layer followed by batch normalization and a ReLU activation function. The final layer of the network maps the refined features back to the original low-resolution SH coefficients.

\subsection{AE-GAN}
Next, the AE-GAN model from \cite{hu2024hrtf} is employed to upsample the denoised SH coefficients. The autoencoder consists of an encoder and a decoder network, where the encoder extracts the latent representation, $z$, of the low-degree SH coefficients, and the decoder reconstructs the high-resolution coefficients. As a refinement, we incorporate channel attention blocks within the residual blocks of the encoder, allowing the model to adaptively focus on important frequency components by dynamically weighting channel-wise features. Additionally, a discriminator is integrated into the model to distinguish between the SH coefficients produced by the generator and those from the real data, ensuring the authenticity of the generated outputs. To further enhance the diversity and sparsity of the generated SH coefficients and improve the realism of individualized HRTFs, we integrate minibatch discrimination \cite{salimans2016improved} into the original discriminator network. This mechanism allows the model to assess not just individual samples but their variations within a batch, promoting higher sparsity of input and greater diversity in the generated outputs.

% a discriminator is integrated to distinguish between real and generated SH coefficients, ensuring authenticity. To further improve the realism of individualized HRTFs, we refine the model by incorporating minibatch discrimination \cite{salimans2016improved} into the discriminator. This enhancement allows the model to capture batch-wise variations, promoting greater diversity and sparsity in the generated SH coefficients.

% \subsection{Data Post-processing}
% Finally, to transform the HRTF data back into the frequency domain from the SH coefficients, \( F_{l}^{m} \), after upsampling, the inverse SHT can be applied, which is defined as
% \begin{equation}
% f(\theta, \phi) = \sum_{l=0}^{\infty} \sum_{m=-l}^{l} F_{l}^{m} Y_{l}^{m}(\theta, \phi)\:.
% \end{equation}

\section{Experimental Setup}
\label{sec:experiments}
\subsection{Data Generation for Training}
% The proposed U-GANet is evaluated on a test set of 41 subjects from the SONICOM HRTF dataset \cite{engel2023sonicom}. Each HRTF in the SONICOM dataset contains a total of 793 different positions for each individual. 
The training and testing data come from the SONICOM HRTF dataset \cite{engel2023sonicom}, employing 793 positions per HRTF.
% A HRTF, which characterizes how sound is filtered by the listener's anatomy before reaching the ears, can be represented in the frequency domain as:
% \begin{equation}
% \text{HRTF}(f, \theta, \phi) = \frac{P(f, \theta, \phi)}{P_0(f)}\:,
% \end{equation}
% where \( P(f, \theta, \phi) \) is the pressure at the listener's ear for a sound source at azimuth \( \theta \) and elevation \( \phi \), and \( P_0(f) \) is the reference pressure.

To simulate real-world HRTF measurement noise, we add both white and pink noise in the time domain to the clean HRTFs from the SONICOM dataset. Noise is added to the left and right sides of the HRTF independently before the impulse responses from the two ears are concatenated together. The noise component, \( N(t) \), which can comprise either of white noise or pink noise, is defined as follows,
\begin{equation}
N_{\text{white}}(t) \sim \mathcal{N}(0, \sigma^2) \:,
\end{equation}
where $N_{\text{white}}(t)$ represents the white noise at time $t$ and $\mathcal{N}(0, \sigma^2)$ indicates that the white noise follows a gaussian distribution with Mean $0$ and variance $\sigma^2$.

\begin{equation}
N_{\text{pink}}(t) = \frac{1}{M} \sum_{i=1}^{M} \text{randn}_i(t)\:,
\end{equation}
where \( M \) is the number of random sources, typically set to 16 in the Voss-McCartney algorithm. Noise is added at the desired signal-to-noise ratio (SNR) using the following approach.
\begin{equation}
\text{HRTF}_{\text{noisy, white}}(t) =  \text{HRTF}(t) + N_{\text{white}}(t)\:,
\end{equation}
where \( \text{HRTF}_{\text{noisy, white}}(t) \) is the white noisy HRTF signal in the time domain and \( \text{HRTF}(t) \) is the original HRTF signal.
\begin{equation} 
\resizebox{0.89\hsize}{!}{$
\text{HRTF}_{\text{noisy, pink}}(t) = \text{HRTF}(t) + \frac{1}{M} \sqrt{\frac{P_{\text{signal}}}{\text{SNR}_{\text{linear}} \cdot P_{\text{noise}}}} \sum_{i=1}^{M} \text{randn}_i(t)
$\;.}
\end{equation}
where \( \text{HRTF}_{\text{noisy, pink}}(t) \) is the pink noisy HRTF signal in the time domain, \( P_{\text{signal}} = \frac{1}{T} \sum_{t=1}^{T} \text{HRTF}(t)^2 \) and \( P_{\text{noise}} = \frac{1}{T} \sum_{t=1}^{T} N(t)^2 \) denote the average power of the original HRTF and noise, respectively. The signal-to-noise ratio in linear scale is given by \( \text{SNR}_{\text{linear}} \), where \( \text{randn}_i(t) \) represents the \( i \)-th random noise source and \( M \) is the total number of noise sources. The noisy HRTFs are then downsampled to generate the low-resolution noisy data for training.

% To achieve a desired signal-to-noise ratio (SNR) in decibels, the noise is scaled according to:

% \[
% \text{Scaling Factor} = \sqrt{\frac{P_{\text{signal}}}{\text{SNR}_{\text{linear}} \times P_{\text{noise}}}}
% \]

% where:
% \begin{itemize}
%     \item \( P_{\text{signal}} = \frac{1}{T} \sum_{t=1}^{T} \text{HRTF}(t)^2 \) is the average power of the original HRTF signal,
%     \item \( P_{\text{noise}} = \frac{1}{T} \sum_{t=1}^{T} N(t)^2 \) is the average power of the noise,
%     \item \( \text{SNR}_{\text{linear}} = 10^{\left(\frac{\text{SNR}_{\text{dB}}}{10}\right)} \) converts the SNR from decibels to a linear scale.
% \end{itemize}

% The scaled noise is then added to the clean HRTFs to obtain the final noisy HRTFs for each channel:

% \[
% \text{HRTF}_{\text{noisy,left}}(t) = \text{HRTF}_{\text{clean,left}}(t) + \left( N(t) \times \text{Scaling Factor} \right)
% \]
% \[
% \text{HRTF}_{\text{noisy,right}}(t) = \text{HRTF}_{\text{clean,right}}(t) + \left( N(t) \times \text{Scaling Factor} \right)
% \]

% The noisy HRTFs are maintained as separate channels for the left and right ears:

% \[
% \text{HRTF}_{\text{noisy}}(t) = \begin{cases}
% \text{HRTF}_{\text{noisy,left}}(t) & \text{(Left Ear)} \\
% \text{HRTF}_{\text{noisy,right}}(t) & \text{(Right Ear)}
% \end{cases}
% \]

% In the frequency domain, these noisy HRTFs can be transformed back into the time domain using the inverse Fourier transform:

% \[
% h(t, \theta, \phi) = \mathcal{F}^{-1}[\text{HRTF}(f, \theta, \phi)]
% \]

\subsection{Training}

\subsubsection{Denoisy U-Net Training}
\label{sec:U-GANet_training}
The model is trained using a combination of loss functions, primarily L1 loss,

\begin{equation}
\mathcal{L}_{L1} = \frac{1}{N} \sum_{i=1}^{N} |\text{SH}_{\text{denoised},i} - \text{SH}_{\text{target},i}|\:,
\end{equation}

where \(\text{SH}_{\text{target},i}\) is the ground truth clean SH coefficient, \(\text{SH}_{\text{denoised},i}\) is the denoised SH coefficient, and \(N\) is the number of coefficients.

Additionally, a cosine similarity loss (CSL) is used to ensure that the angular similarity between the denoised SH coefficients and target coefficients is maximized. The cosine similarity loss is defined as,
\begin{equation}
\mathcal{L}_{\text{cos}} = 1 - \frac{\sum_{i=1}^{N} \text{SH}_{\text{denoised},i} \cdot \text{SH}_{\text{target},i}}{\sqrt{\sum_{i=1}^{N} \text{SH}_{\text{denoised},i}^2} \cdot \sqrt{\sum_{i=1}^{N} \text{SH}_{\text{target},i}^2}}\:.
    \label{eq:cos loss}
\end{equation}
% These two loss functions are combined to supervise U-Net training.
% Total loss function for the U-Net can be expressed as follows: 
% \begin{equation}
%     \mathcal{L}_{total} = \lambda\mathcal{L}_{L1} + \mathcal{L}_{\text{cos}}, 
% \end{equation}
% where $\lambda$ is a weight applied to the L1 loss term, ensuring that it will not become too large and disrupt the training process.

\subsubsection{AE-GAN Training}
The discriminator is trained via supervised learning, utilizing both generated and real HRTF data, and aims to guide the autoencoder to produce high-fidelity results. In this study, we further extend the application of AE-GAN by expanding the range of sparcity levels to include 4 points and 3 points, thus demonstrating the model's robustness and scalability across a broader spectrum of resolutions.

\subsubsection{HRTF-DUNet Training}
For end-to-end training and joint optimization, we employ Cascaded Backpropagation, enabling seamless gradient flow between the U-Net denoiser and AE-GAN upsampler. This refinement ensures that U-Net optimally denoises SH coefficients to support AE-GAN’s upsampling, enhancing reconstruction accuracy, especially under extreme sparsity conditions.

\subsection{Baselines}
The performance of the proposed approach is compared against four baselines: AE-GAN without DUNet, barycentric interpolation, SH interpolation, and non-individual HRTF selection. The AE-GAN approach is presented in \cite{hu2024hrtf}. Barycentric interpolation estimates unknown values by computing weighted averages of known points using three barycentric coordinates. SH interpolation, widely used for HRTF upsampling~\cite{arend2021assessing}, projects HRTF data onto SH for smooth spatial representation. An alternative to personalized HRTF modeling selects the closest match from a database. Following~\cite{hogg2024hrtf}, Selection-1 represents the most `generic' HRTF, while Selection-2 is the most `distinct'.

\subsection{Evaluation Metrics}
Three metrics are used for evaluating the performance.

\subsubsection{Interaural level difference (ILD)}
\label{sec:ILD}
The ILD represents the interaural level difference, which is the difference in sound pressure level between the two ears for a given frequency $f_b$, number of spatial locations $N$, total number of frequency bins $B$ and direction  $x_n$, calculated by,
% \begin{equation}
% \text{ILD}(f, \theta, \phi) = 20 \log_{10}\left(\frac{P_{\text{left}}(f, \theta, \phi)}{P_{\text{right}}(f, \theta, \phi)}\right)\:,
%     \label{eq:ILD}
% \end{equation}
\begin{equation}
    \resizebox{0.89\hsize}{!}{$
    \begin{aligned}
    \text{ILD} = \frac{1}{N}\sum_{n=1}^{N}\frac{1}{B}\sum_{b=1}^{B}
    &\Bigg|\Bigg(20\log_{10}\frac{|H_\text{LR}^\text{Left}(f_{b},x_{n})|}{|H_\text{LR}^\text{Right}(f_{b},x_{n})|}\Bigg) \\ &- \Bigg(20\log_{10}\frac{|H_\text{DN}^\text{Left}(f_{b},x_{n})|}{|H_\text{DN}^\text{Right}(f_{b},x_{n})|}\Bigg)\Bigg|
    \end{aligned}$\;,}
    \label{eq:ILD}
\end{equation}
The terms $|H^\text{Left}(f_b, x_n)|$ and $|H^\text{Right}(f_b, x_n)|$ denote the magnitude responses for the left and right ears, respectively. Similarly, $|H_\text{LR}(f_b, x_n)|$ and $|H_\text{DN}(f_b, x_n)|$ represent the magnitude responses for the low-resolution and denoised \text{HRTF} sets. 
% Here, $B$ is the total number of frequency bins in the \ac{HRTF}, $N$ is the number of spatial locations, $f_b$ refers to the frequency, and $x_n$ specifies the location.
% where \( P_{\text{left}}(f, \theta, \phi) \) represents the sound pressure at the left ear for a given frequency \( f \) and direction \( (\theta, \phi) \) and \( P_{\text{right}}(f, \theta, \phi) \) is the sound pressure at the right ear for the same frequency and direction.

\subsubsection{Interaural time difference (ITD)}
\label{sec:ITD}
The ITD, quantifies the arrival time gap of a sound wave between the left and right ears for the same frequency and direction, given by,
% \begin{equation}
% \text{ITD}(f, \theta, \phi) = \frac{\Delta t_{\text{left-right}}}{T}\:,
%     \label{eq:ITD}
% \end{equation}
\begin{equation}
    \resizebox{0.89\hsize}{!}{$
    \begin{aligned}
    \text{ITD} = \frac{1}{N}\sum_{n=1}^{N}\frac{1}{B}\sum_{b=1}^{B}
    &\Bigg|\Bigg(\frac{\phi_\text{LR}^\text{Left}(f_{b},x_{n}) - \phi_\text{LR}^\text{Right}(f_{b},x_{n})}{2\pi f_b}\Bigg) \\ &- \Bigg(\frac{\phi_\text{DN}^\text{Left}(f_{b},x_{n}) - \phi_\text{DN}^\text{Right}(f_{b},x_{n})}{2\pi f_b}\Bigg)\Bigg|
    \end{aligned}$\;,}
    \label{eq:ITD}
\end{equation}

where $\phi_\text{LR}^\text{Left}(f_b, x_n)$ and $\phi_\text{LR}^\text{Right}(f_b, x_n)$ represent the phase responses of the low-resolution HRTF for the left and right ears. Similarly, $\phi_\text{DN}^\text{Left}(f_b, x_n)$ and $\phi_\text{DN}^\text{Right}(f_b, x_n)$ correspond to the denoised HRTF phase responses.

\subsubsection{Log-spectral distortion (LSD)}
\label{sec:LSD}
The LSD \cite{gutierrez2022interaural} is an evaluation metric utilized to assess the quality of a synthesized audio signal relative to a reference audio signal. In this context, LSD is employed to evaluate the denoising and upsampling performance of HRTFs using the proposed HRTF-DUNet framework. The LSD loss quantifies this comparison by evaluating the discrepancy between the target magnitude spectrum $H_{\text{HR}}$ and the generated spectrum $H_{\text{G}}$. This computation can be expressed in the following way,
\begin{equation}
\resizebox{0.89\hsize}{!}{$
    \text{LSD} = \frac{1}{N}\sum^N_{n=1}\sqrt{\frac{1}{W}\sum^W_{w=1}\left( 20\text{log}_{10} \frac{|  H_{\textbf{HR}}(f_w, x_n)   |}{ | H_{\text{G}}(f_w, x_n) | } \right)^2} 
\:,$}  
    \label{eq:LSD}
\end{equation}
where N represents the overall count of positions, and $x_n$ corresponds to a specific position.

\section{Experimental Results}
Two experiments were performed to evaluate the newly proposed HRTF-DUNet model. These experiments utilised 41 test subjects (HRTFs) not seen in training from the SONICOM dataset, where white noise was added at an SNR of 5dB.
\subsection{Denoising Evaluation}
\begin{figure}[t]
    \centering
    \subfloat{
    \includegraphics[width=0.95\linewidth]{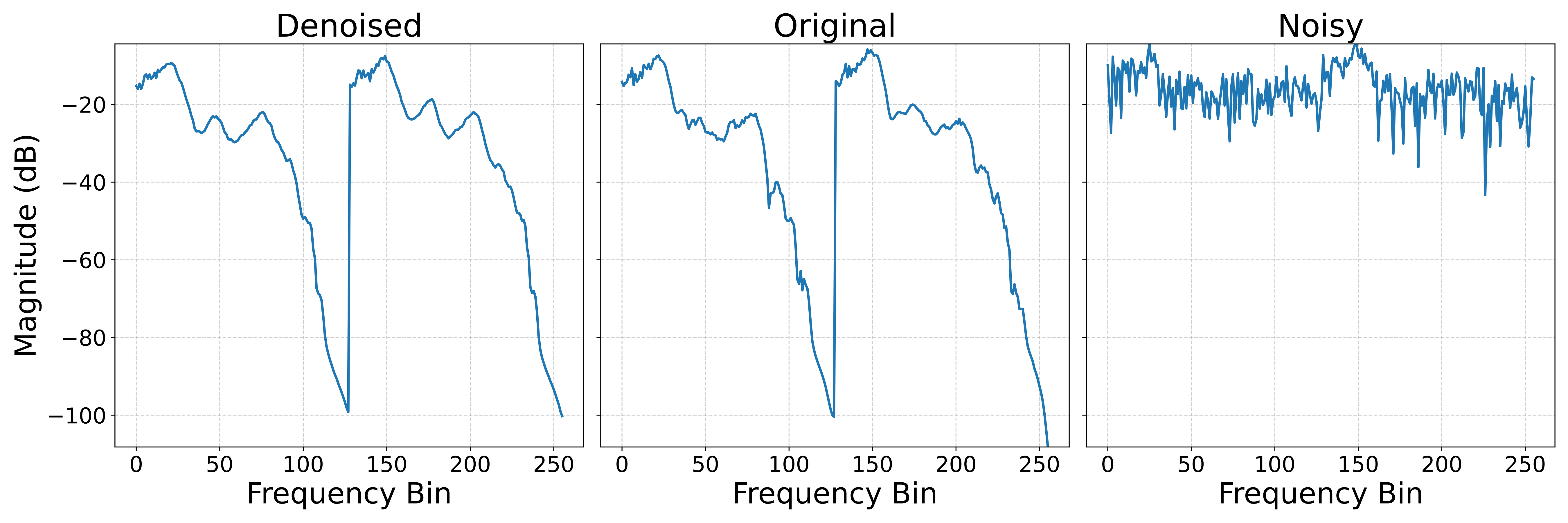}
    } 
    
    \subfloat{%
    \includegraphics[width=0.95\linewidth]{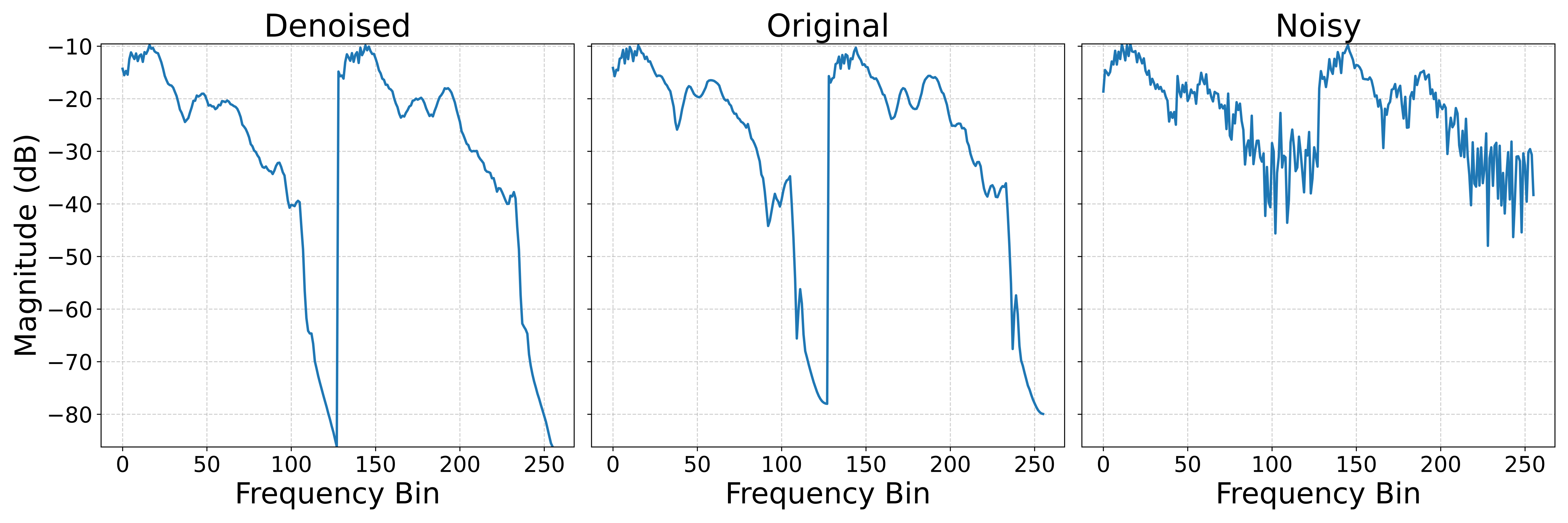}
    } 
    \vspace{-0.15cm}
    \caption{Two illustrative examples (top and bottom) showcasing the HRTF Denoisy U-Net's performance on two different subjects at the same measurement location, with additive white Gaussian noise applied at an SNR of 5 dB.}
    \label{fig:denoising}
    \vspace{-0.35cm}
\end{figure}

The proposed Denoisy U-Net model for HRTF denoising is evaluated against three baseline methods: Spectral Subtraction, Wavelet and Kalman Filtering. Table \ref{tab:denoise} presents the performance comparison across three evaluation metrics: CSL (detailed in Section \ref{sec:U-GANet_training}), ILD (outlined in Section \ref{sec:ILD}), and ITD (described in Section \ref{sec:ITD}). Fig. \ref{fig:denoising} shows results of the HRTF Denoisy U-Net.

The results demonstrate that the HRTF Denoisy U-Net model outperforms the baselines across all metrics. For example, CSL, which measures the similarity between the denoised and target HRTFs, is significantly lower for the HRTF Denoisy U-Net model (0.007), indicating a higher degree of similarity and better denoising capability. Additionally, the U-Net model shows superior performance in preserving ILDs and ITDs, with the lowest deviations of 19.757 and 1.301, respectively. These results indicate that the denoising process effectively preserves critical spatial cues that are contained within the HRTFs and which are needed for realistic, immersive audio.

\begin{table}[!tb]
\centering
\renewcommand{\arraystretch}{1.25}
\setlength{\tabcolsep}{2.0pt}
\caption{A comparison of HRTF Denoisy U-Net and baselines with different evaluation metrics (`best' result highlighted).}
\label{tab:denoise}
\resizebox{0.7\linewidth}{!}{%
\begin{tabular}{|c|c|c|c|c|c|}
\hhline{-~---}
\textbf{Method} & & \textbf{CSL} & \textbf{ILDs} & \textbf{ITDs}  \\
\hhline{=~===}
\textbf{HRTF Denoisy U-Net} & & \cellcolor{blue!25}0.007 & \cellcolor{blue!25}19.757 & \cellcolor{blue!25}1.301  \\
\hhline{-~---}
\textbf{Wavelet Filtering with dB7} & & 0.283 & 24.591 & 2.783 \\
\hhline{-~---}
\textbf{Wavelet Filtering with Gaus3} & & 0.213 & 22.178 & 2.846 \\
\hhline{-~---}
\textbf{High-Pass Spectral Subtraction} & & 0.339 & 23.936 & 2.946 \\
\hhline{-~---}
\textbf{Kalman Filtering} & & 0.206 & 20.491 & 2.152 \\
\hline
\end{tabular}}
\vspace{-0.1cm}
\end{table}

\subsection{LSD Evaluation}
\begin{figure}[!tb]
\centering
\includegraphics[width=0.95\linewidth]{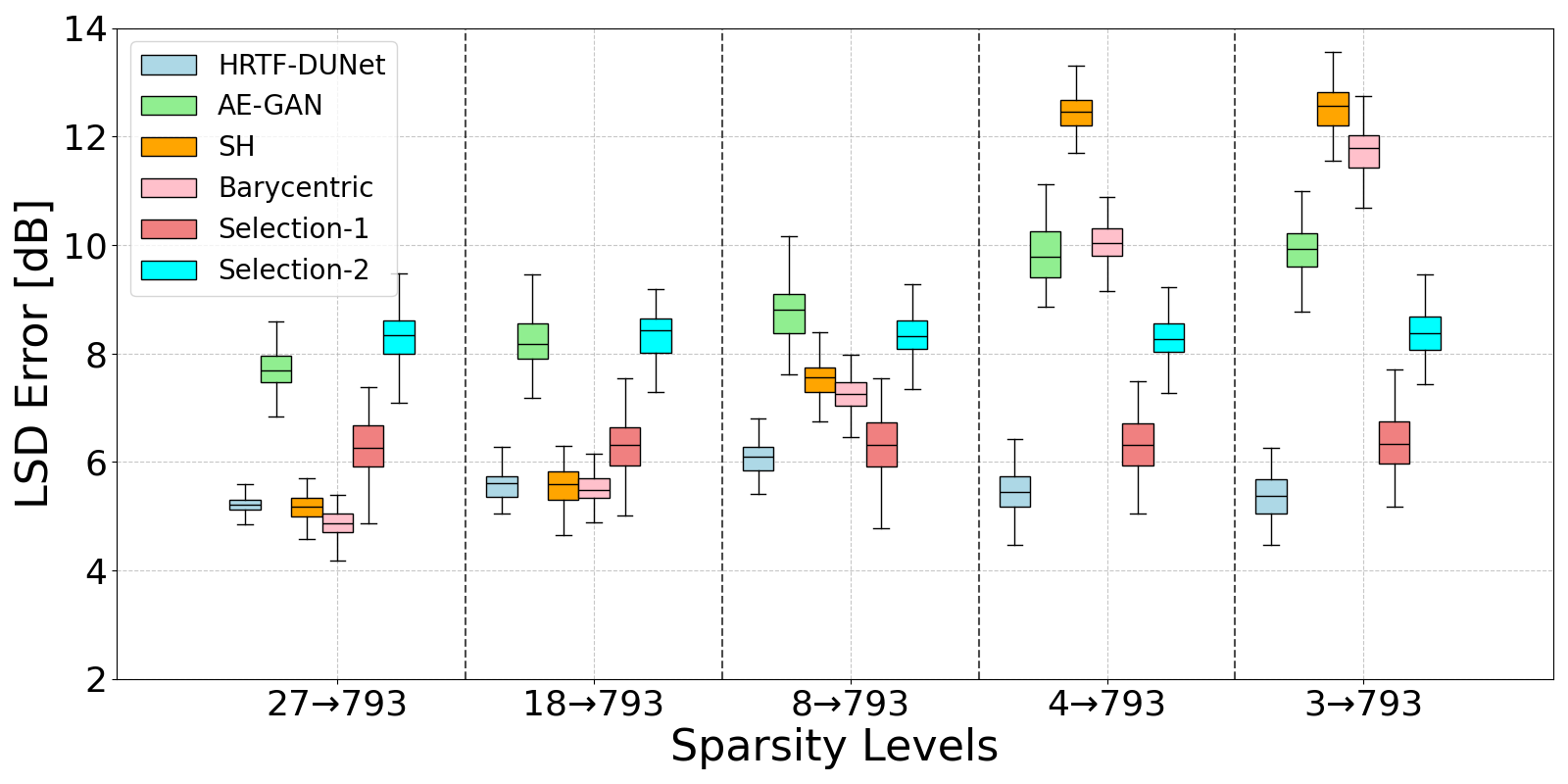}
\vspace{-0.15cm}
\caption{\label{fig:lsd_box}Log-spectral distortion (LSD) error comparison.}
\vspace{-0.35cm}
\end{figure}

Second, the HRTF-DUNet model is evaluated using the LSD metric (see Section \ref{sec:LSD}), averaging the LSD across all measurement positions. Table \ref{tab:lsd result} presents the results for 41 noisy test subjects, with a visualization in Fig. \ref{fig:lsd_box}.

The results show that the HRTF-DUNet model consistently achieves a lower LSD error at sparcity levels $4\rightarrow793$ and $3\rightarrow793$, where it significantly outperforms other baselines. This suggests that HRTF-DUNet effectively denoises and learns patterns in noisy HRTF features, even in extremely sparse conditions. Unlike traditional interpolation methods, which rely on a predefined spatial structure, the deep-learning-based approach can generalize from the available data and reconstruct a more accurate HRTF. Barycentric and SH interpolation yield higher LSD errors at extreme sparsity levels as their geometric assumptions break down. However, with sufficient initial points, they perform well by leveraging spatial smoothness and structured mathematical formulations for accurate interpolation. AE-GAN struggles at high sparsity levels due to insufficient spatial information and noise in the limited initial HRTF points, preventing effective feature learning. However, as sparsity decreases, AE-GAN benefits from more input data, enabling better feature extraction and improved high-resolution HRTF reconstruction. The HRTF selection approach performs poorly, with LSD errors of 6.31 and 8.33 for Selection-1 and Selection-2, respectively. This reinforces the limitation of non-individualized HRTFs, emphasizing the need for personalization to achieve realistic virtual audio.

% Barycentric and SH interpolation, on the other hand, tend to produce larger LSD errors at higher sparsity level, which corresponds to scenarios where the number of initial HRTF data points is very sparse. This is to be expected since their geometric assumptions just break apart. However, they perform well when a sufficient number of initial points are available, as they effectively leverage spatial smoothness and rely on well-defined mathematical structures for accurate interpolation.

% AE-GAN struggles at high sparsity levels because it fails to learn meaningful features from the extremely limited and noisy initial HRTF points. With too few initial points, the autoencoder lacks sufficient spatial information to construct a reliable latent representation, leading to inaccurate upsampling. Additionally, the discriminator can't provide strong supervision in such sparse conditions, further hindering the model’s ability to generalize. However, as sparsity level decreases, AE-GAN starts to perform better. With more input data, the encoder can extract meaningful spatial features, allowing the decoder to generate more accurate high-resolution HRTF coefficients.

% The HRTF selection approach, on the other hand, performs poorly and yields LSD errors of 6.31 and 8.33 for Selection-1 and Selection-2, respectively. These results highlight the known fact that non-individualized HRTFs generally result in high LSD errors and underscore the current desire for personalized HRTF modeling that is often needed to create realistic virtual audio.

\begin{table}[!tb]
\centering
\renewcommand{\arraystretch}{1.25}
\setlength{\tabcolsep}{2.0pt}
\caption{A comparison of the mean LSD error (Standard Deviation) for different sparsity levels (`best' performance highlighted).}
\label{tab:lsd result}
\resizebox{0.95\linewidth}{!}{%
\begin{tabular}{|c|c|c|c|c|c|c|}
\hhline{-~-----}
\multirow{2}{*}{\textbf{Method}} & & \multicolumn{5}{c|}{\textbf{Upsampling [No. intial $\rightarrow$ No. upsampled]}} \\ 
\hhline{~~-----}
 & & \textbf{27 $\rightarrow$ 793} & \textbf{18 $\rightarrow$ 793} & \textbf{8 $\rightarrow$ 793} & \textbf{4 $\rightarrow$ 793} & \textbf{3 $\rightarrow$ 793} \\
\hhline{=~=====}
\textbf{HRTF-DUNet} & & 5.23 (0.19) & 5.58 (0.28) & \cellcolor{blue!25}6.06 (0.32) & \cellcolor{blue!25}5.43 (0.45) & \cellcolor{blue!25}5.41(0.41) \\
\hhline{-~-----}
\textbf{AE-GAN} & & 7.74 (0.41) & 8.20 (0.49) & 8.76 (0.55) &  9.70 (0.56) & 9.89 (0.51) \\
\hhline{-~-----}
\textbf{SH} & & 5.12 (0.27) & 5.54 (0.31) & 7.54 (0.37) & 12.46 (0.39) & 12.41 (0.44) \\
\hhline{-~-----}
\textbf{Barycentric} & & \cellcolor{blue!25}4.89 (0.24) & \cellcolor{blue!25}5.46 (0.27) & 7.22 (0.35) & 10.07 (0.43) & 11.69 (0.47) \\
\hhline{-~-----}
\textbf{Selection-1} & & \multicolumn{5}{c|}{6.31 (0.59)} \\
\hhline{-~-----}
\textbf{Selection-2} & & \multicolumn{5}{c|}{8.33 (0.47)} \\
\hhline{-~-----}
\end{tabular}}
\vspace{-0.1cm}
\end{table}

% \subsection{Perceptual Evaluation}

\section{CONCLUSION AND FUTURE WORK}
\label{sec:conclusion}

This paper introduces a novel framework using the HRTF-DUNet model for simultaneous HRTF denoising and upsampling, simplifying the measurement process for personalised HRTFs. To the best of our knowledge, this is the first work to address the problem of HRTF denoising, demonstrating its feasibility and advantages in improving HRTF quality. The proposed method outperforms other approaches by effectively denoising and upsampling three measurement points with 5 dB of additive white noise into a high-resolution, clean HRTF.

This work serves as a proof of concept, showing that denoising is not only possible but also beneficial in the context of HRTF upsampling. In future work, we will extend beyond simulated white and pink noise, applying this method to more realistic simulated and recorded noise to better reflect real-world environments.

Additionally, challenges still remain in measuring HRTFs in uncontrolled environments, particularly problems with reverberation and frequency range limitations. These issues arise due to the reverberation present when measuring in untreated rooms and the frequency limitations imposed by the speakers used for recording the HRTFs. Addressing these challenges is an ongoing focus of our research. We also plan, going forward, to conduct perceptual evaluations using computational auditory models and listening tests to further validate the model’s effectiveness in practical immersive audio applications.

\bibliographystyle{IEEEbib}
\bibliography{axdstrings,refs}

\end{document}